\documentclass{article}
\usepackage{here}
\usepackage{graphicx}
\evensidemargin
0.cm
\def\beq{\begin{equation}}
\def\eeq{\end{equation}}
\def\bea{\begin{eqnarray}}
\def\eea{\end{eqnarray}}
\def\bq{\begin{quote}}
\def\eq{\end{quote}}

\def\nnb{\nonumber}
\def\ga{\left(}
\def\dr{\right)}

\def\rar{\rightarrow}
\def\nnb{\nonumber}
\def\la{\langle}
\def\ra{\rangle}
\def\nin{\noindent}
\def\ba{\begin{array}}
\def\ea{\end{array}}

\begin{document}
\date{}
\title{
\vskip-20mm
\begin{flushright}{\small
MPI-PhT-2001-20\\
PM/01-34}
\end{flushright}
\vskip 30mm
{\bf Hints on the power corrections 
from \\ current
correlators
in $x$-space}}
\author{S. Narison$^{\rm a}$ and  V.I.
Zakharov$^{\rm b}$
\\
\\
$^{\rm
a}$ {\small\it Laboratoire de Physique
Math\'ematique,
Universit\'e
de Montpellier 2,}\\
{\small\it Place
Eug\`ene Bataillon
34095 -
Montpellier Cedex 05, France}\\
{\small\it Email: narison@lpm.univ-montp2.fr}\\
$^{\rm b}$
{\small\it Max-Planck Institut f\"ur Physik, F\"ohringer Ring
6,
80805
M\"unchen, Germany}\\
{\small\it Email:
xxz@mppmu.mpg.de}
}
\sloppy
\maketitle

\begin{abstract}\noindent
We
consider an interpretation of the recent lattice data
on the
current-current correlators in the $x$-space.
The data indicate
rather
striking difference between
(axial)vector and (pseudo)scalar
channels
which goes beyond
the predictions of the standard
non-perturbative models.
We argue that if the
difference is to be
explained by power corrections,
there is a unique
choice of the form
of the correction. We discuss the
emerging picture
of the power
corrections.
\end{abstract}
\setcounter{page}{1}

\pagestyle{plain}
\section{Introduction}
\nin
We shall be concerned here
with the current-current correlators in the coordinate
space:
\beq\label{twopoint}
\Pi(x)= \la
0|J(x)J^\dagger(0)|0\ra~,
\eeq
in
case of
  the $(V\pm A)$ and (pseudo)scalar
   currents:
\bea
J_\mu^{(V\pm
A)}&=&\bar q_i\gamma_\mu (1\pm \gamma_5)q_j~,\nnb\\
J^{S\pm P}&=&[(m_i-
m_j)\pm (m_i+m_j)]\bar
q_i(1\pm\gamma_5)q_j~,
\eea
where $q_{i,j}$ and
$m_{i,j}$ are the
quark fields and masses.
The two-point functions
(\ref{twopoint})
obey a dispersion representation:
\beq
\Pi(x)=\frac{1}{4\pi^2}\int_0^\infty
dt~\frac{\sqrt{t}}{x}K_1(x\sqrt{t})
~{\rm
Im}\Pi(t)~,
\eeq
where Im$\Pi(t)$ is related to the current induced cross section and $K_1(z)$
is the modified Bessel function, which behaves
for small $z$
as:
\beq
K(z\rar 0)\simeq  \frac{1}{z}+\frac{z}{2}\ln
z~.
\eeq
In the
limit $x\to 0$, $\Pi(x)$ coincides with the free-field
correlator and the
main theoretical issue is how the asymptotic
freedom gets violated at
intermediate $x$.\\
  From pure theoretical point of view, the use of
the $x$-space is no better
than the use of the momentum space, which is
the  traditional tool of the
QCD sum rules \cite{SVZ,SNB}. Each
representation
has its own
advantages and inconveniences (for a recent discussion see
\cite{SHEVCHENKO}). The $x$-space
approach is motivated and described in detail in Ref.
\cite{SHURYAK2}. In particular, the current correlators
(\ref{twopoint})
are measured in the most direct
way on the lattice.
The importance of the lattice measurements
\cite{CHU,DEGRAND} is
that they allow to study the correlation functions for 
currents with  various
quantum numbers, while direct
experimental
information is confined to only vector and axial-vector currents
\cite{ALEPH,OPAL}. The well-known $\tau$-decay data were widely
used for
theoretical analyses both in the $Q$- and $x$-spaces ( see, e.g.,
\cite{BNP,SNVA,ioffe,SHURYAK} ).
Most recently,  new lattice data on the
$S,P$ channels were  obtained \cite{DEGRAND}.
The most interesting observation is that in the $S+P$
channel there are
noticeable deviations
from the instanton liquid
model \cite{SHURYAK2}
while in the $V\pm A$
channels the agreement of
the existing
data with this model
is quite good
\cite{SHURYAK,DEGRAND}. \\
Such deviations were in fact predicted in Ref.
\cite{CNZ} where
unconventional
quadratic corrections, $\sim 1/Q^2$ were introduced.
The primary aim
of the present note is to perform a more detailed
comparison of the lattice data
with the model of Ref. \cite{CNZ}.
We, indeed, find further support for
the quadratic corrections. However, the
overall picture is far from being complete
and we are trying to
analyze
the data in a more generic way. The central assumption
is  that the
violations of the parton model for the correlators at
moderate $x$
are due to power-like corrections.
\section
{Current-current correlators}
For the sake of
completeness, we begin
with a summary of theoretical expressions
for the current
correlators, both in the $Q-$ and $x-$spaces.
We will focus on the $(V\pm A)$ and $(S\pm P)$ channels since the recent lattice
data \cite{DEGRAND} refer to these channels.
In case of $(V\pm A)$ currents the
correlator is defined as:
\beq\label{odin}
\Pi_{\mu\nu}(q)=i\int
d^4x~e^{iqx}\la TJ_{\mu}(x)J_{\nu}(0)^{\dagger}\ra
=
(q_{\mu}q_{\nu}-g_{\mu\nu}q^2)\Pi(q^2)~,
\eeq
where $-q^2\equiv Q^2>0$
in the Euclidean space-time.
For the sake of
definiteness we fix
the
flavor structure of the light-quark current
$J_{\mu}$
as:
\beq
J_\mu^{V\pm A}=\bar{u}\gamma_\mu(1\pm\gamma_5)
d~.
\eeq
In the
chiral limit one has in the $(V+A)$ case (see,
e.g.,
\cite{SVZ,SNB}):
\beq\label{sum}
\Pi^{V+A}(Q^2)~=~{1\over
2\pi^2}\Bigg{\{}-\ga
1+{\alpha_s\over \pi}\dr\ln {Q^2\over\nu^2}
-{\alpha_s\over
\pi}{\lambda^2\over Q^2} +{\pi\over
3}{\la\alpha_s(G^a_{\mu\nu})^2\ra\over
Q^4}+{256\pi^3\over
81}{\alpha_s\la\bar{q}q\ra^2\over
Q^6}\Bigg{\}}~.
\eeq
The corresponding  relation for the $(V-A)$ case
reads as:
\beq\label{difference}
\Pi^{V-A}(Q^2)~=
-\frac{64\pi}{9}{\alpha_s\la\bar{q}q\ra^2\over
Q^6}~,
\eeq
In the x-space the same
correlators, upon dividing by $\Pi^{V+A}_{pert}$
where $\Pi^{V+A}_{pert}$
stands for
the perturbative correlator, are obtained by
applying the equations
collected for convenience in the Table
\ref{fourier}. 
\begin{table}[H]
\begin{center}
\setlength{\tabcolsep}{3.pc}
\caption{Some
useful Fourier
transforms}
\label{fourier}
\begin{tabular}[H]{ll}
\hline
&\\
$Q$-space&$x$-spa
ce\\
&\\
\hline
&\\
$Q^2\ln
Q^2$&$\frac{8}{\pi^2}\frac{1}{x^6}$\\
$\ln
Q^2$&$-\frac{1}{\pi^2}\frac{1}{x^4}$\\
$\frac{1}{Q^2}$&$\frac{1}{4\pi^2}\frac{1}{x^2}$\\
$\frac{1}{Q^2}\ln Q^2$&$-\frac{1}{4\pi^2}\frac{1}{x^2}\ln^2x^2$\\
$\frac{1}{Q^4}$&$-\frac{1}{16\pi^2}\ln{x^2}$\\
$\frac{1}{Q^4}\ln Q^2$&$\frac{1}{64\pi^2}\ln^2x^2$\\
$\frac{1}{Q
^6}$&$\frac{1}{8\times
16\pi^2}x^2\ln{x^2}$\\
$\frac{1}{Q^6}\ln Q^2$&$-\frac{1}{496\pi^2}x^2\ln^2x^2$\\
$\frac{1}{Q^8}$&$-\frac{1}{8\times
16\times
24\pi^2}x^4\ln{x^2}$\\
$\frac{1}{Q^8}\ln Q^2$&$\frac{1}{496\times 24\pi^2}x^4\ln^2x^2$\\
&\\
\hline
\end{tabular}
\end{center}
\end{table}
\nin
In particular,
\beq
{\Pi^{V+A}\over
\Pi^{V+A}_{pert}}~\rightarrow~1-{\alpha_s\over
4\pi}\lambda^2\cdot
x^2~-~
{\pi\over 48}\la\alpha_s(G^a_{\mu\nu})^2\ra
x^4\ln
x^2~+~{2\pi^3\over
81}\alpha_s\la\bar{q}q\ra^2x^6\ln
x^2~.
\eeq
Note that $\ln x^2 $ is negative since we start from small $x$.
An  important technical
point is that on the lattice one measures the trace
over the Lorentz
indices $\mu,\nu$, see Eq~(\ref{odin}).
In the $Q$-space this is equivalent
to considering $Q^2\cdot\Pi(Q^2)$
instead of $\Pi(Q^2)$.
The
$x$-transform of the $Q^2\cdot\Pi(Q^2)$
is given
by:
\beq\label{vplusa}
{Q^2\cdot\Pi^{V+A}\over Q^2\cdot
\Pi^{V+A}_{pert}}~\rightarrow~
1~-~{\pi\over
96}\la\alpha_s(G^a_{\mu\nu})^2\ra x^4~
+~{2\pi^3\over
81}\alpha_s\la\bar{q}q\ra^2x^6\ln x^2.
\eeq
Next, we will concentrate on the currents having
the
quantum
numbers of the pion and
of $a_0(980)$-meson.
The correlator
of two
pseudoscalar currents is defined
as
\beq\label{pion}
\Pi^P(Q^2)~\equiv~ i\int
d^4x~e^{iqx}\la
T\{J^{\pi}(x)J^{\pi}(0)\}\ra~,
\eeq
where
\beq
J^P~=~i(m_u+m_d)\bar{u}\gamma_5d~,
\eeq
In the momentum space, to represent the result in terms of the running
coupling, and  masses it is more convenient to consider the second derivative in 
$Q^2$ of $\Pi^P(Q^2)$
defined in Eq. (\ref{pion}), which obeys an homogeneous RGE:
\beq\label{pionb}
\frac{\partial^2\Pi^P}{(\partial Q^2)^2}~=~{3\over
8\pi^2}\frac{(\bar
m_u+\bar m_d)^2}{Q^2}\Bigg{\{}1+\frac{11}{3}{\bar\alpha_s\over
\pi} \\-{4\alpha_s\over
\pi}\frac{\lambda^2}{ Q^2 }
+2{\pi\over 3}{\la\alpha_s(G_{\mu\nu}^a)^2\ra\over Q^4}
+2\cdot 3{896\pi^3\over
81}{\bar\alpha_s\la\overline{\bar{q}q}\ra^2\over
Q^6}\Bigg{\}}~.
\eeq
Here, the
standard OPE terms can be found in
\cite{SVZ,SNB,BECCHI} while the
gluon-mass correction
was introduced first
in \cite{CNZ} \footnote{We assume that $\alpha_s \lambda^2$ does not run like
$\la\alpha_s(G_{\mu\nu}^a)^2\ra$.}. 

In what
follows, we shall work with the
appropriate
ratio where the pure
perturbative corrections are absorbed
into the
overall normalization and
concentrate on the power corrections
assuming that these
corrections are responsible
for the observed rather sharp
variations of the correlation functions.
Thus, in the x-space we
have for  the pion channel:
\beq\label{pionx}
{\Pi^{P}\over
\Pi^{P}_{pert}} ~\rightarrow~
1-{\alpha_s\over 2\pi}\lambda^2x^2
+{\pi\over
96}\la\alpha_s(G_{\mu\nu}^a)^2\ra x^4
-{7\pi^3\over
81}\alpha_s\la\bar{q}q\ra^2x^6\ln x^2~.
\eeq
Note that the coefficient in front of the last
term in Eq. (\ref{pionx}) differs
both in the absolute value and sign from the
corresponding expression in \cite{SHURYAK2}.
The channel which is crucial
for our analysis is the $(S+P)$. In this
channel:
\bea\label{splusp}
R_{P+S}\equiv \frac{1}{2}\ga{\Pi^{P}\over
\Pi^{P}_{pert}}+{\Pi^{S}\over \Pi^{S}_{pert}}\dr
~\rightarrow~
1-{\alpha_s\over
2\pi}\lambda^2x^2+{\pi\over
96}\la\alpha_s(G_{\mu\nu}^a)^2\ra
x^4+{4\pi^3\over
81}\alpha_s\la\bar{q}q\ra^2x^6\ln
x^2~.
\eea
This
expression concludes the summary of the power corrections to the
current
correlators.
\section{Quadratic power
corrections.}
One of the central points of the present note is that
there
are no $\lambda^2$ corrections to the $V\pm A$ correlators
as can be seen
in Eq
(\ref{vplusa}).
On the other hand, these terms
are present in the case of the $(S\pm P)$ channels
as can be seen in
Eq (\ref{splusp}). There is no such asymmetry in the
$Q$-space, see Eqs
(\ref{sum}), (\ref{difference})
and
(\ref{pion}). Thus, the $(V\pm A)$ correlator, as
measured on the lattice,
are $\lambda^2$-term blind \footnote{Note that a reversed case is
well known. Namely,
there are certain terms which are seen in
the small-$x$ expansion and
are not seen in the large-$Q$ expansion
\cite{SHURYAK2}.}~! Thus, we
are coming to {\it a
kind of a theorem}.
Namely:

\nin
If one assumes:\\
(1) that  the $V\pm A$ channels are
described by the instanton
liquid model while in the $S+ P$ channel there
are considerable
deviations from this model
   (as the lattice data seem to
strongly
indicate \cite{DEGRAND}) \\
and\\
(2) that this difference is due to
some power corrections,\\
then \\
the power corrections can be uniquely
identified as the gluon-mass corrections
(see terms proportional to
$\lambda^2$).

\nin
Note that the $\lambda^2$ corrections are singled out
for two reasons:
   First, since
taking the trace over the Lorentz indices
$\mu,\nu$ corresponds to
multiplying by
$Q^2$ as can be seen in the
discussion of Eq. (\ref{vplusa}) and
it is only a $1/Q^2$  correction which
can become a polynomial as
a result of
multiplying $\Pi(Q^2)$ by $Q^2$.
Second, in the $Q$-space there should be  no log factor in front of
$1/Q^2$, $1/Q^2\cdot
\ln Q^2$.
These two conditions are satisfied in case of $(V\pm A)$
currents
and are not fulfilled in the $(S\pm P)$ channels. The difference
between the channels is that in the
latter case the $\lambda^2$ correction
is present in the imaginary part of the
$\Pi^{S,P}(Q^2)$ \cite{CNZ}.
\\
Thus, if one retains only the
$\lambda^2$ corrections, then, there are
no violations
of the parton
(perturbative) picture in the $(V\pm A)$
channels for the correlator
measured in \cite{DEGRAND} while the violations
are present in the
$(S\pm P)$ channels. \\
Of course, this limiting case
is not necessarily describing the reality
and we proceed to quantitative
fits to the data \cite{DEGRAND}.
\section{Analysis of the data}
\begin{figure}[htb]
\begin{center}
\includegraphics[width=13cm]{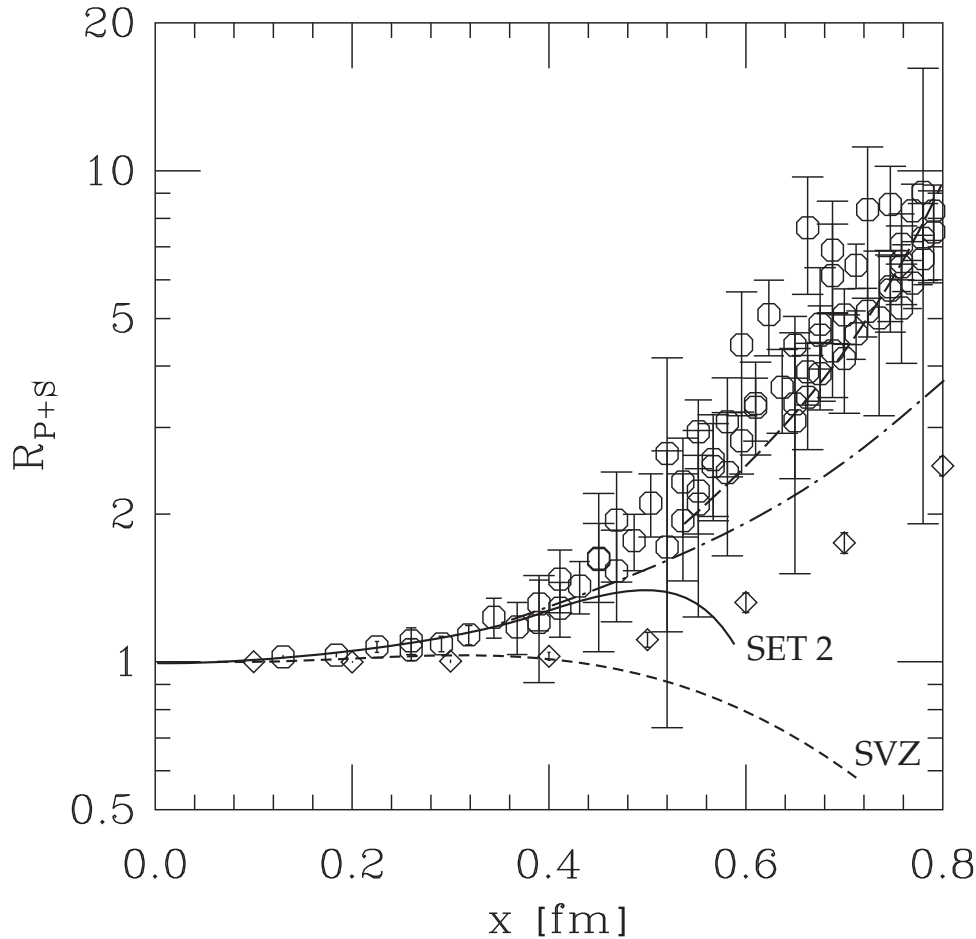}
\caption{$S+P$ channel: comparison  of the lattice data from \cite{DEGRAND} with the 
OPE
predictions for
the two Sets of QCD condensate values given in Table
2. The dot-dashed curve is the prediction for SET 3 where the
contribution of the $x^2$-term has been added to SET 2. The bold dashed curve
is SET 3 + a fitted value of the $D=8$ condensate contributions.
The diamond curve is the prediction from the instanton liquid
model of \cite{SHURYAK}. }
\end{center}
\end{figure}
\nin
In Fig. 1
we confront the OPE predictions with the lattice data
on the $(S+P)$
channel obtained in \cite{DEGRAND}. The choice of the
$(S+P)$ channel is
motivated by the fact the single instanton
contribution cancels from this
channel \footnote{The instanton contribution
is not large also in the
$(V\pm A)$ channels. However the $\lambda^2$ terms are
canceled from these
channels, see the discussion in section 3,
and we cannot add anything to
the analysis of Ref. \cite{SHURYAK,DEGRAND}.}
and it was predicted in Ref.
\cite{CNZ} that the $\lambda^2$ correction
will be manifested in this
channel.
The theoretical curves in Fig. 1 correspond to two sets of
values
of the
condensates given in Table 2. The first set (SET 1) corresponds
to
the standard SVZ values of the
gluon and four-quark condensate, the latter
being obtained using the
vacuum saturation assumption. The second set (SET
2), corresponds to
the values of the
condensates obtained in
\cite{SNGE},
where the value of the gluon condensate is two times the
SVZ value and
the
four-quark condensate exhibits a violation of the vacuum saturation,
first obtained from
$e^+e^-$ data in \cite{LAUNER}. In SET 3, one also
accounts the
presence of the new
1/$Q^2$-term first advocated in \cite{CNZ}
and fitted from $e^+e^-$ data in
\cite{SN2,SNGE}
\footnote{A
common difficulty encountered in determining the quadratic corrections
is
that they usually compete with the standard perturbative
radiative
corrections. In
\cite{SN2,SNGE}, a suitable choice of the sum
rules (e.g. ratio of
moments) has been used
such that the perturbative
radiative corrections are eliminated to
leading order and
the contribution
of the quadratic term becomes optimal.
Moreover, the quadratic corrections
corresponding to $\lambda^2\approx - 0.5 GeV^2$
do not affect in a
significant way the
determination  of $\alpha_s$ from
$\tau$-decay as has
been explicitly shown in \cite{CNZ}. On the
contrary, the  quadratic
term
appears to decrease very slightly $\alpha_s$ from the $\tau$-decay and
bring it closer to
the
world average value at $M_Z$. In Ref. \cite{ioffe}
bounds were obtained
on the value of $\lambda^2$ from the sum rules which
have large perturbative terms.
This turned possible due to a particular
fixation of the
of the perturbative terms in the complex $q^2$ plane.
In
particular, the common use of the running coupling
would affect the
procedure strongly \cite{ioffe}.}. Note also that
in numerical fits we put
$\ln x^2=-1$, the same, as, say, in
Ref.\cite{SHURYAK}.
\begin{table}[H]
\begin{center}
\setlength{\tabcolsep}{2.pc}
\caption{Different
parameters used in the analysis of the $S+P$ data
in units of GeV$^d$
($d$
is the dimension of the
operator).}
\begin{tabular}[H]{llll}
\hline
&\\
Sources&
$\la\alpha_s G^2\ra$&$\alpha_s\la\bar\psi\psi\ra
^2$&$(\alpha_s/\pi)\lambda^2$\\
&\\
\hline
&\\
SET 1 (SVZ)
\cite{SVZ}&0.04& $0.25^6$&0\\
&\\
SET 2 \cite{SNGE}&0.07&$5.8\times
10^{-4}$&$0$\\
&\\
SET 3 \cite{SNGE,CNZ}&0.07&$5.8\times
10^{-4}$&$-0.12$\\
&\\
\hline
\end{tabular}
\end{center}
\end{table}
\nin
The analysis indicates that a much better fit of the lattice data for
the $S+P$ channel at
moderate values of
$x$ is achieved after the inclusion of the
$1/Q^2$, or $x^2$ quadratic
correction. A caveat is that we account only
for the power corrections, not pure perturbative contributions.
The reason is that the lattice data, in their
present status,
do not give any clear indication of the perturbative
contributions.
Note also that
the data cannot discriminate between the
values of the dimension four
and six condensates
entering in SET 1 and SET 2 as the effects of these two condensates tend
to compensate each other for
the choice
$\ln x^2\approx
-1$. The agreement of the OPE with the lattice data at larger values 
of $x$ can be obtained
by the inclusion of the $D=8$ condensate with a size $+(x/0.58)^8$ 
where we have used $\ln^2
x^2\simeq 1\simeq -\ln x^2$. This value can be compared with the one 
$+(3395/30855168)\la\alpha_s G^2\ra^2 x^8\approx (x/1.2)^8$, which one would obtain
from  the evaluation of these
contributions in \cite{BROAD} and where a modified factorization of 
the gluon condensates
proposed in \cite{LATORRE} has been used. 
\begin{figure}[htb]
\begin{center}
\includegraphics[width=13cm]{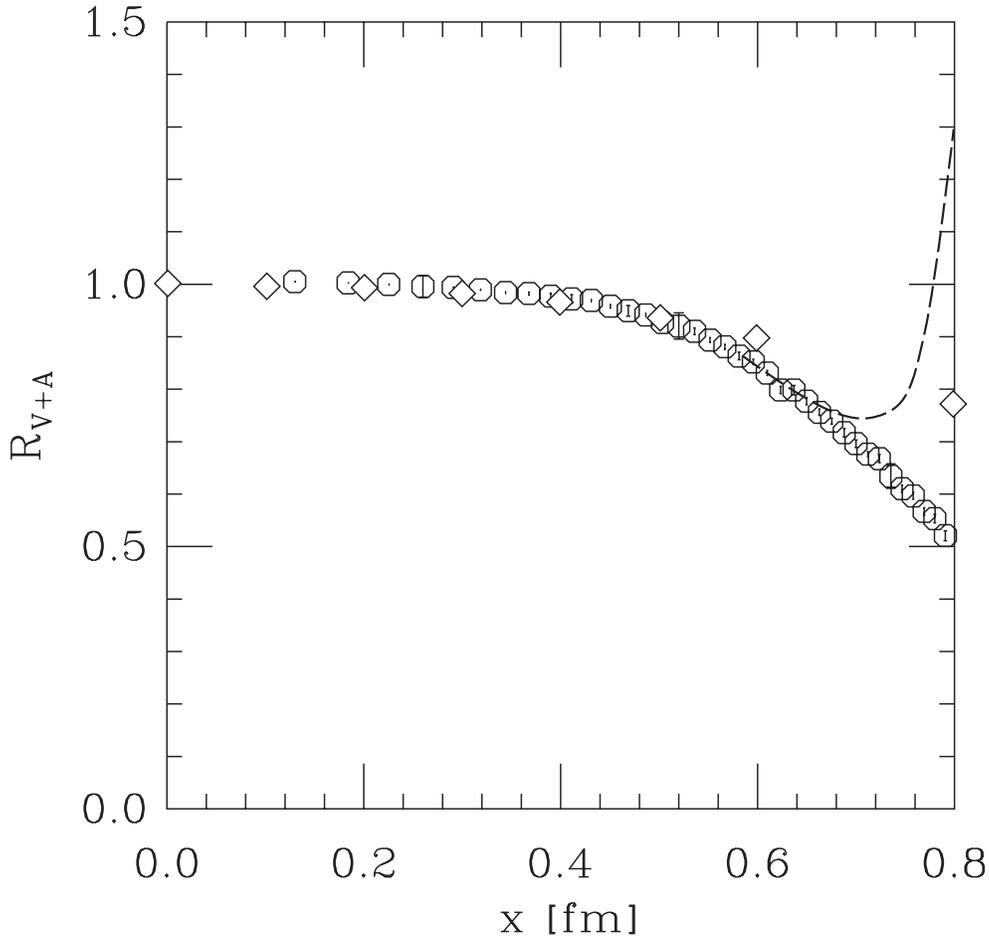}
\caption{$V+A$ channel: comparison  of the lattice data from \cite{DEGRAND} with the 
OPE
predictions for
the SET 3 QCD condensates values given in Table
2 including a fitted value of the  $D=8$ contributions. The diamond curve is the prediction
from the instanton liquid model of \cite{SHURYAK}. }
\end{center}
\end{figure}
\nin
For completion, we show in 
Fig. 2, a fit of the
lattice data in the V+A channel using SET 3 values of the gluon and 
quark condensates and quadratic term plus a 
$D=8$ contribution with the strength
$(x/0.7)^8$ to be compared with the one $\la\alpha_s 
G^2\ra^2x^8/3428352\approx(x/2.5)^8$ which
one would obtain using the results in
\cite{BROAD,LATORRE}. Both fits in Figs 1 and 2 might indicate that 
the vacuum saturation can
be strongly violated for higher dimension condensates, a feature 
already encountered from
different analysis of the $\tau$ and $e^+e^-$ data 
\cite{SNGE,RAFA2,ALEPH,OPAL}. Therefore, we
would also expect analogous large deviations in the V-A channel.
\section{Discussions. Two-step QCD}
While evaluating the emerging picture of the power corrections,
one
should face the possibility that  the standard OPE (see, e.g.,
\cite{SVZ,SNB})
is valid only at very short distances. What is even more
important,
the mass scale where higher terms in the OPE become numerically comparable
to the lowest ones
is not necessarily the scale associated with
the resonances but could be considerably higher.
There is
accumulating
evidence to support such a view:\\
(1) A direct comparison of the OPE with the lattice data in the $(V-A)$
channel
demonstrates that the convergence radius of the OPE is no larger
than 0.3 fm
\cite{SHURYAK,DEGRAND}. \\
(2) Within the instanton liquid model \cite{SHURYAK2},
the distance between
instantons is a few times larger than the size of the instantons.
On the
physical grounds, the OPE applies at distances smaller than the instanton
size
while the resonance properties are rather related to distance between
the instantons
(if encoded in the model at all). Respectively, neither the
lattice data nor the predictions
of the instanton liquid model exhibit any
irregularity at the convergence radius of the OPE \cite{SHURYAK}. \\
(3) Within the monopole-dominated-vacuum model the two scales
are even
more pronounced numerically: the monopole radius is about 0.06 fm while
the
distance between the monopoles is about 0.5 fm \cite{BORNYAKOV}.\\
(4) If one replaces the local condensates of the standard OPE
by their non-local counterparts (for a review see
\cite{DOSCH}), then the effect of the
non-locality is strong already at
$( 0.1\div 0.3)$ fm \cite{HOFMANN}.

If, indeed, the validity of the standard OPE derived within the fundamental
QCD
is shrunk to very short distances, then improving  within the
standard OPE fits to
the data obtained at presently available lattices might not
be a proper criterion
for selecting the right model.

Instead, there emerges a picture according to which the non-standard
quadratic
power corrections dominate the presently available
``intermediate distances '' of
about $(0.1\div 0.5)$ fm. 
There are two pieces of evidence of the quadratic
corrections
dominating over the whole range of intermediate distances
indicated above:\\
(1) non-perturbative contribution to the heavy quark potential is linear at
all the distances
$r> 0.1$ fm (for discussion see \cite{AKHOURY}).\\
(2) Instanton density, as function of the instanton size $\rho$ is
reproduced at all the
distances $\rho>0.1$ fm \cite{SHURYAK4}.\\
Moreover,\\
(3) Introduction of the tachyonic gluon mass explains in a simple
and unified way existence of the strong channel dependence of the
scales of the violation of the asymptotic freedom \cite{CNZ}.\\
To this list we add now a new observation:\\
(4) Instanton-liquid model plus the $\lambda^2$ correction gives a
reasonable fit
in the $(S+P)$ channel at distances $(0.1\div 0.5)$ fm.

As for the status of the $\lambda^2$ correction it is most solid
within the effective Higgs-like theories which are common 
within the monopole mechanism of confinement. Indeed,
in the presence of a magnetically charged (effective) scalar field 
the symmetry of the theory is $SU(3)_{color}\times U(1)_{magnetic}
$ \cite{CHERNODUB}. Upon the spontaneous breaking of the magnetic
$U(1)$ the gauge boson acquires a non-vanishing mass
and its mass squared is the only parameter of dimension $d=2$
consistent with the symmetry.
Moreover, in exchanges between (color) charged particles the
gauge-boson mass appears to be tachyonic mass 
as was demonstrated on the $U(1)$ example in Refs. \cite{GUBAREV,SHEVCHENKO}.
Detailed analysis of various power corrections within the
Higgs-like models can be found in Refs. \cite{GUBAREV,SHURYAK4,SHEVCHENKO}.
Moreover, if the monopole size is indeed as small as indicated above,
then the effective Higgs-like theories  
can apply at all the distances $\sim(0.1\div 0.5)$ fm. 

To summarize, the $\lambda^2$ corrections introduced in
\cite{CNZ} rather
drastically improve agreement
of the theoretical predictions for the
current correlator in the $(S+P)$ channel
with the lattice data  (without
affecting the other channels measured), in the moderate $x$-region, and which
can be extended to higher $x$-values by including higher dimension condensates.
The main uncertainty of the
analysis is due to neglect of the pure perturbative
corrections, not
detected so far on the lattice.
Further checks could be provided by
measuring the current correlators in the other channels
discussed in \cite{CNZ}. The success of the fits with $\lambda^2$
corrections included can be understood within the effective theories.

\section*{Acknowledgements}
The authors are
grateful to Th. De Grand, F.V. Gubarev, E.V. Shuryak
for useful discussions
and communications. S. Narison wishes to thank W-Y. Pauchy 
Hwang for the hospitality at
CosPA-NTU (Taipei), where some part of this work has been done.
V. Zakharov would like to acknowledge gratefully the hospitality
of the Centre National de la Recherche Scientifique (CNRS) during his
stay at the University of Montpellier when this was begun.  

\end{document}